\documentstyle[aps,psfig]{revtex} 		

\def\be{\begin{equation}}
\def\ee{\end{equation}}
\def\beqa{\begin{eqnarray}}
\def\eeqa{\end{eqnarray}}

\def\Psian{\tilde{\Psi}({\bf r},t)}

\def\br{{\bf r}}

\begin{document}
\title{Josephson effect between Bose condensates}

\newcommand{\beq}{\begin{equation}}
\newcommand{\eeq}{\end{equation}}

\author{
F. Sols
}
\address{
Departamento de F\'{\i}sica Te\'orica
de la Materia Condensada,
Universidad Aut\'onoma de Madrid, E-28049 Madrid, Spain 
}
\maketitle

\begin{abstract}
\end{abstract}


\section{INTRODUCTION}

Since its theoretical prediction in 1962 
\cite{jose62}, the Josephson effect has played a major role in the 
physics and technology of superconductors \cite{baro82}. The physics of the
Josephson
effect becomes manifest when a 
weak link is created ({\it e.g.} by tunneling or through a point contact)
between 
two systems which have undergone some type of gauge symmetry 
breaking. In fact, the Josephson effect has also been observed in
superfluids \cite{aven85}, and, because of the profound analogies, its
observation in the recently achieved \cite{ande95,brad95,davi95,mewe96}
Bose condensed atomic gases is 
generally expected. A system whose gauge symmetry has been 
spontaneously broken can be described by an order parameter that 
behaves in many respects like a macroscopic wave function $\Psi ({\bf 
r})$. In the simplest cases, the order parameter reduces to a complex scalar,
$\Psi=\sqrt{\rho}e^{i\phi}$, where $\rho({\bf r})$ is the ``superfluid 
density" and $\phi({\bf r})$ is the phase. For Bose-Einstein condensates of
dilute alkali gases, 
$\Psi(\br)$ is the 
wave function of the macroscopically occupied one-atom state. 

In his epoch making paper \cite{jose62}, Josephson predicted that, between
two 
weakly connected superconductors of phases $\varphi_1$ and $\varphi_2$, a 
non-dissipative particle (Cooper pair) current flows between them 
whose value is
\beq
I(\varphi)=I_c \sin\varphi,
\eeq
where $I_c$ is the critical current and $\varphi=\varphi_2-
\varphi_1$ is the relative phase [$\phi({\bf r})$ is assumed to be
approximately uniform within each superconductor]. He also predicted that,
in the 
presence of a nonzero chemical potential difference $\mu=\mu_2-
\mu_1$, the relative phase rotates as
\beq
\dot{\varphi} =-\mu/\hbar
\eeq
The Josephson relations can be obtained from very general 
considerations and apply to any pair of weakly linked systems that can be
described 
by macroscopic wave functions \cite{feyn63}. They can be obtained 
as the equations of motion of the ``pendulum Hamiltonian"
\begin{equation}
\label{pendulum}
H(\varphi,N)= E_J(1-\cos \varphi)+\frac{1}{2}E_CN^2,
\end{equation}
where $E_J=\hbar I_c$ is the Josephson coupling energy, $N=(N_2-
N_1)/2$ is the number of transferred particles, and $E_C 
\equiv \partial \mu/\partial N$ is the capacitive energy due to
interactions. In the absence of external constraints, $\mu=E_C N$.
In a superconducting link, the critical electric current is 
$2eI_c$ and the capacitive (or ``charging" ) energy is 
$E_C=(2e)^2/2C$, where $C$ is the electrostatic capacitance. For 
trapped BEC's, $E_C$ can be obtained to a good accuracy from the 
Thomas-Fermi (TF) calculation \cite{baym96} of the chemical potential  [see
eq. (\ref{EC})].

When both $E_C$ and $k_BT$ are $\ll E_J$, the Josephson pendulum Hamiltonian
can be approximated as a harmonic oscillator,
\beq
\label{HO}
H(\varphi,N) \simeq \frac{1}{2} E_J \varphi^2 + \frac{1}{2}E_C N^2,
\eeq
whose frequency 
\beq
\label{WJP}
\omega_{JP}=\sqrt{E_JE_C}/\hbar
\eeq
is called the Josephson plasma frequency.

Given the intimate relationship between eqs. (1-2) and eq. (3), a 
derivation of the pendulum Hamiltonian (in particular of the 
$-\cos \varphi$ dependence of the energy on the relative phase) may be
regarded as equivalent to a derivation of the Josephson effect. 
At low temperatures, the collective dynamics of an inhomogeneous BEC is
well described by the Gross-Pitaevskii (GP) Hamiltonian \cite{pita61}
\be
\label{GPH}
H[\Psi,\Psi^*] = \int d{\bf r} \left( \frac{\hbar^2}{2m}
|\nabla \Psi(\br)|^2
+
V_{\rm ext} (\br)|\Psi(\br)|^2 +\frac{g}{2}|\Psi(\br)|^4 
\right),
\ee
where $g=4\pi\hbar^2a/m$ ($a$ is the $s$-wave scattering length). 
Neglecting depletion \cite{gior97}, the normalization 
can be taken as $\int d{\bf r} \rho=N_T$, being $N_T$ the
total number of atoms. The extremum condition $\delta(H-\mu_0N_T)/\delta
\Psi^*=0$ (where $\mu_0$ is the equilibrium chemical potential) leads to
the well known GP equation \cite{pita61}.
 
\section{PHASE RIGIDITY AND BROKEN GAUGE SYMMETRY}

The assumption of a uniform phase within each Bose condensate is implicit
in most discussions of the Josephson effect. It seems therefore convenient
to check explicitly that this is a physically meaningful approximation. To
prove it, let us estimate the energy of a
one-radian
fluctuation of the phase across the condensate near equilibrium
in a single spherical harmonic well
[see eq. (\ref{GPH})],
\beq
\label{EFl}
\frac{\hbar^2}{2m}\int d{\bf r}\rho |\nabla 
\phi|^2\simeq \frac{\hbar^2}{2m}\frac{N_T}{4R^2}
\simeq 0.7 N_T^{3/5}  \frac{ \hbar
\omega_0}{2},
\eeq
where $R=a_0(15aN_T/a_0)^{1/5}$ is the cloud radius
estimated within the
TF approximation \cite{baym96}, $\omega_0$ is the 
harmonic
oscillator frequency of the well, and 
$a_0=(\hbar/m\omega_0)^{1/2}$
is the oscillator length. For the last approximate equality we 
have
used typical parameters $a=5$ nm and $a_0=10^{-4}$ cm. The
characteristic temperature of such an oscillation can be as big as
10 $\mu$K \cite{mewe96}. Since temperatures as low as 100 nK can 
be reached in current BEC setups, it is acceptable to view the 
phase of a condensate within one well as internally frozen and 
rigid. It is clear from (\ref{EFl}) that spatial fluctuations of 
the phase are energetically more costly the higher the value of 
the condensate density $\rho$. For this same reason, in the 
interestitial region between two wells, where
$\rho({\bf r})$ decreases appreciably, spatial phase variations 
require less energy and are thus easier to activate at low 
temperatures. Here lies the essence of the 
low energy Josephson dynamics. At
low temperatures, the picture emerges of two wells, 
each with an internally uniform phase, 
whose relative phase 
\beq
\label{phase}
\varphi=\int_1^2 d{\bf l} \cdot \nabla \phi,
\eeq
induced by local variations of $\phi({\bf r}$) in the 
interstitial region, is the only active dynamical variable. 

From the theory of phase transitions, we know that some
type of ``rigidity" is generally exhibited in the ordered phase below the
critical temperature \cite{ande84}. 
In the case of superconductors, superfluids, and Bose-Einstein 
condensates (BEC's), one may speak of ``phase rigidity" because, given 
an electromagnetic gauge (or its equivalent for superfluids and 
BEC's), the phase is determined everywhere once it has been fixed 
at a given point. Alternatively, since gauge and phase are 
intimately connected in quantum mechanics, we can say that, 
given a choice of phase, we have lost the freedom to choose the 
gauge, hence the term ``gauge symmetry breaking".

Long range phase coherence can be destroyed by thermal or quantum
fluctuations. A simple picture of how this occurs can be developed by
analyzing the Josephson Hamiltonian (\ref{pendulum}), which may be viewed
as a two-site reduction of the more general energy functional (\ref{GPH}).
In equilibrium, good phase coherence between sites 1 and 2 is obtained in
the harmonic limit (\ref{HO}). Then it is appropriate to speak of global
phase coherence or rigidity in the combined double BEC system. Thermal
fluctuations randomize the phase for $k_BT \agt E_J$. To see how quantum
fluctuations can destroy the two-site phase coherence, we note that, since
$N$ and $\varphi$ are canonically conjugate variables, $N$ can be
represented as $N=i\partial/\partial\varphi$ in a quantum description. As a
consequence, the capacitive term in (\ref{pendulum}) plays the role of the
kinetic energy for the phase variable. At zero temperature, quantum
fluctuations can destroy phase rigidity between sites 1 and 2 if the
interactions are sufficiently strong ($E_C \gg E_J$). In this limit, the
relative particle number becomes a good quantum number [see eq.
(\ref{pendulum})] and the two condensates behave as essentially
independent. Because of the weak connection, it is much easier to destroy
the relative phase coherence between condensates 1 and 2, than the internal
phase rigidity within each condensate. We have already proved that internal
coherence cannot be destroyed by thermal fluctuations at the experimental
temperatures. On the other hand, long range coherence within each
condensate is preserved from destruction by quantum fluctuations thanks to
the three-dimensional character of the problem, as stated by the
Mermin-Wagner theorem in quantum statistical mechanics.

\section{EXTERNAL JOSEPHSON EFFECT}

\subsection{Semiclassical description}

Let us assume that $V_{\rm ext}(\br)$ is such that our system has the
structure of two weakly connected condensates $1$ and $2$.
We assume that the condensates are confined within spherical
harmonic wells of the same frequency $\omega_0$.
First we wish to analyse the semiclassical dynamics. 
Then, $N$ and $\varphi$ can be treated as 
simultaneously well-defined, and we may write for the wave function the 
ansatz \cite{zapa98}
\be
\label{ansatz}
\Psian \sim \Psi_1({\bf r}; N_T/2 -
N(t))+e^{i\varphi(t)}\Psi_2({\bf r}; N_T/2 + N(t)),
\ee
where $\Psi_i({\bf r}; n)$ is the (real) equilibrium wave function 
for
the isolated well $i$ containing $n$ bosons. It is straightforward 
to
show that, to lowest order in the overlap integrals $\int 
\Psi_1
\Psi_2$, the energy functional for $\tilde{\Psi}$ takes the form
\be
\label{RedEn}
H(\varphi,N)
\simeq E_B
(N) + E_J(N)\left(1 - \cos \varphi \right),
\ee
where $E_B(N)$ is the bulk energy of the two isolated wells
with $N$ transferred atoms, and $E_J(N)$ is
the Josephson coupling energy.

In the TF limit \cite{baym96}, the bulk energy $E_B(N)$ is mostly due to
interactions, and it may be expanded as
\be
\label{EBulk}
E_B(N)\simeq E_{B}(0)+\mu_0'(\frac{N_T}{2})N^2 +
\frac{1}{12} \mu_0'''(\frac{N_T}{2})N^4.
\ee
Since $\mu_0 \sim N_T^{2/5}$ 
\cite{baym96},
the ratio between the third and second terms in the expansion is 
$0.32 \,
N^2/N_T^2$, which means that the last term can be neglected in a
wide range of situations. To avoid complications stemming from
possible resonances between Josephson oscillations (see below) and
intrawell excitations, we require $\mu_0(N_T/2+N)-\mu_0(N_T/2-N)\ll 
\hbar \omega_0$, where we use the result that the first normal
mode of a spherical well lies approximately at $\hbar \omega_0$ 
above
the ground state \cite{stri96,edwa96}. This condition is
realized when $N/N_T\ll 4.6 N_T^{-2/5}$ for typical parameters. 
This upper bound is of order $2-10$\% for $N_T \sim 10^4-10^6$.

\subsection{Josephson coupling energy in one dimension}

The expression for the coupling energy $E_J(N)$ in terms of
$\Psi_1$ and
$\Psi_2$ implicit in 
(\ref{RedEn}) 
is rather complicated and 
difficult to handle, so it is desirable to find a simpler expression.
Let us assume that the $(1-\cos\varphi)$ dependence for the energy of a
Josephson link has already been proved by {\it e.g.} the arguments in the
previous subsection, and focus on the particular case $\varphi \ll 1$, when
$\Psi(\br)$ is very close to the ground state wave function $\Psi_0(\br)$
with $\varphi=0$.
In such a case, the integrand $|\nabla\Psi|^2$ in (\ref{GPH}) is mostly due
to phase variation. In one dimension, we can then write
\be
\label{identity}
E[\phi]\equiv\frac{\hbar^2}{2m} \int_1^2 dx \rho(x)\left|\phi'(x)\right|^2  
\simeq \frac{1}{2}E_J\varphi^2,
\ee
where the phase term in (\ref{pendulum}) has been approximated quadratically,
and where the integration extends between two points beyond which the phase
is practically uniform. 

From (1), we know that a supercurrent must flow between the two condensates
whose value is
\beq
\label{Iphi}
I \simeq (E_J/\hbar)\varphi.
\eeq
On the other hand, particle number conservation requires
\beq
\label{Iuni}
I=(\hbar/m)\rho(x)\phi'(x)={\rm cnst.}
\eeq
Inserting the resulting expression for $\phi'(x)$ into (\ref{identity}),
and using
 (\ref{Iphi}), one obtains a functional expression for the Josephson
coupling energy in terms of the ground state superfluid density:
\beq
\label{functional}
E_J = \frac{\hbar^2}{m}\left[\int_1^2 \frac{dx}{\rho_0(x)} \right]^{-1},
\eeq
where we have used the fact that, for $\varphi \ll 1$,
$\rho\simeq \rho_0\equiv|\Psi_0|^2$.

\subsection{Three dimensions: Electrostatic analogy}

We assume again that the $(1-\cos\varphi)$ dependence has already been
proved and consider the case $\varphi \ll 1$.
Analogously to (\ref{identity}), we can write 
\be
\label{electro}
E[\phi]\equiv\frac{\hbar^2}{2m} \int_{\rm ext} d{\bf r}
\rho(\br)|\nabla\phi(\br)|^2  
\simeq \frac{1}{2}E_J\varphi^2,
\ee
where the integration extends over the region exterior to the
condensates (the results are quite independent on the precise 
location
of the condensate borders). The phase within the
condensates is assumed to be uniform. 
The only 
way for
eq. (\ref{electro}) to have such a dependence on the total phase
difference $\varphi$ and not on the details of $\nabla\varphi$ is 
that the condition $\delta E[\phi]/\delta
\phi(\br)=(\hbar^2/2m) \nabla \cdot(\rho\nabla\phi)=0$ is satisfied,
something which is guaranteed by current conservation
[$(\hbar/m)\rho\nabla\phi$ is the particle current density]. The alert
reader may have noted that, except for
trivial factors, this is the electrostatic equation for the 
electric
displacement vector ${\bf D} \equiv
\rho \nabla \varphi$ in a medium with a nonuniform dielectric 
constant
$\rho(\br)$. Boundary conditions for $\phi(\br)$ are given by its value at 
the borders of each condensate ($\varphi_1$ and $\varphi_2$), 
which act as conductors in this analogy.
We have a system of two conductors held at a potential difference
$\varphi$ and a dielectric medium surrounding them. Then 
$E[\phi]$ is
essentially the energy of this capacitor. Potential theory tells 
us
that $E[\phi] \sim \varphi^2$, with $E_J$ in (\ref{electro}) playing the
role of the mutual capacitance of the two conductors in the presence of
such a dielectric.

Standard variational arguments
can be invoked to prove that,
for parallel plate boundary conditions \cite{zapa98},
\be
\label{EJCond2}
\int\frac{dxdy}{\int_1^2 dz \rho(x,y,z)^{-1}} \le 
\frac{m E_J}{\hbar^2} 
\le \left[ \int_1^2\frac{dz}{\int dxdy
\rho(x,y,z)}\right]^{-1} 
\ee
(here $z$ is the longitudinal coordinate connecting the centers of the two
wells, and $x,y$ are the transverse coordinates).
The lower bound is obtained by removing the positive term
$(\partial\phi/\partial x)^2+(\partial\varphi/\partial y)^2$ 
from the
energy
functional (\ref{electro}), while the upper bound is derived by taking
$\varphi$ independent of $x,y$. 
If the equilibrium density can be factorized in 
the region controlling the capacitance (a not very restrictive condition in
practice; see above), we can write
$\rho_0(\br)=f(x,y)g(z)$ and the two bounds become identical. Taking
$\br=0$ at the middle point of the double well configuration, and choosing
$x=y=0$ to be the line connecting the two well centers, we can finally
write \cite{zapa98}
\beq
\label{EJCondensator}
E_J= 
A \frac{\hbar^2}{m}\left[ \int_1^2\frac{dz}{\rho_0(0,0,z)}\right]^{-1},
\eeq
being $A=f(0,0)^{-1}\int dx dy f(x,y)$
an effective area. The factorization condition is much more general that it
might appear at first sight, since it only has to be satisfied in the
interstitial region near the top of the barrier giving the dominant
contribution to $E_J$. There $V_{\rm ext}(\br)$ has a saddle-point and thus
it can be decoupled in longitudinal and transverse coordinates. Since
interactions are negligible in the bottle neck region connecting the two
wells, we conclude that the wave function must be factorizable. 

\subsection{WKB calculation}

In order to proceed further, we need an estimate of 
$\rho_0=|\Psi_0|^2$
in the region of interest. Here we reproduce sketchily the WKB calculation
of ref. \cite{zapa98}.
For two 
identical wells in
equilibrium, the
ground state wave function is symmetric in $z$. 
Focussing on the $x=y=0$ line giving the dominant contribution, we write
\be
\label{psiwbk}
\Psi_0(z)=\frac{B}{\sqrt{p(z)}}\cosh\left[\frac{1}{\hbar}\int_0^zdz'
p(z')\right]
\ee
where $p(z)=[2m(V_{\rm ext}(0,0,z)-\mu_0)]^{1/2}$ and $B$ is a
constant to be determined from the work of ref. \cite{dalf96}.
Introducing (\ref{psiwbk}) into
(\ref{EJCondensator}) we obtain
\be
E_J \simeq \frac{\hbar A |B|^2}{m}
\left[2\tanh\left(\frac{S}{2}\right)\right]^{-1},
\ee
where $S\equiv\int_1^2p(z)dz/\hbar$ is the dimensionless instanton action.
One finally gets \cite{zapa98}
\be
\label{EJ2}
E_J\simeq \frac{e^{-S}}{\tanh (S/2)}
\left(\frac{N_T}{2}\right)^{1/3}\left(\frac{15a}{a_0}\right)^{-2/3}
\frac{\hbar
\omega_0}{2}.
\ee
The $N_T^{1/3}$ dependence can be understood qualitatively 
by noting \cite{zapa98,dalf96} that $A\sim R^{2/3}$ and $B\sim R^{1/2}$.
Since, in the TF approximation \cite{baym96}, $R\sim N_T^{1/5}$, one
obtains $E_J\sim N_T^{1/3}$.
Knowing that the critical temperature satisfies $k_B
T_c \simeq N_T^{1/3}\hbar\omega_0$ \cite{gior96}, we may rewrite (\ref{EJ2})
as (for large $S$ and typical parameters)
\be
\label{EJ3}
E_J \sim k_B T_c e^{-S}
\ee
which is reminiscent of the Ambegaokar-Baratoff formula for superconductors
\cite{baro82} if one notes that $e^{-S}$ is the transmission amplitude for
a particle to traverse the barrier.

\subsection{Josephson plasma frequency}

From (3) and (\ref{EBulk}), we have $E_C=2\mu_0'(N_T/2)$, which, in the TF
approximation \cite{baym96} reads
\be
\label{EC}
E_C\simeq \frac{4}{5}\left(\frac{N_T}{2}\right)^{-3/5}
\left(\frac{15a}{a_0}\right)^{2/5}\frac{\hbar \omega_0}{2}.
\ee
From (\ref{WJP}), (\ref{EJ2}) and (\ref{EC}), we obtain
\be
\label{JosPlas}
\omega_{JP} \simeq 
\left(\frac{2a_0}{15aN_T}\right)^{2/15}
\frac{e^{-S/2}}{\sqrt{\tanh (S/2)}} \,\, \omega_0,
\ee
Since, usually, $\omega_0/2\pi \simeq 10-100$ Hz, 
we
conclude 
that $\omega_{JP}/2\pi \alt 10$ Hz.

The ratio $E_J/E_C$ is a good measure of the classical character 
of the relative phase $\varphi$.
From (\ref{EJ2}) and (\ref{EC}), we find
\be
\label{ratio}
\frac{E_J}{E_C}\simeq 
\left(\frac{2a_0}{15aN_T}\right)^{1/15}
\frac{N_Ta_0}{a}\frac{e^{-S}}{\tanh(S/2)}.
\ee
By varying $S$, the system can be driven from the classical 
regime
($E_J\gg E_C$) to the strong quantum limit ($E_J\ll E_C$). 
However, in
the latter case, quantum effects are only important if  we 
operate
at ultralow temperatures $k_BT \alt E_C$. From (\ref{ratio}), one concludes
that, for typical numbers, $S$ must be roughly $\agt 10$ for quantum
fluctuations to be important.

\section{EFFECT OF DISSIPATION}

So far, we have described the dynamics of a
conservative system. In real life, however, one should expect
a certain amount of 
damping. The
most obvious source of such damping
is the incoherent exchange of normal atoms, and a quantitative
discussion requires a generalization of our results to
nonzero temperature. It is possible to perform a qualitative exploration of
the effect of damping based on a simple transport model \cite{zapa98}. We
can expect thermally excited normal atoms to follow the usual laws of
dissipative currents. Specifically, they must give an Ohmic
contribution to the current,
\be
\label{Ohmic}
I_n=- G \mu,
\ee
where $\mu$ is a spontaneous or induced chemical potential difference (for 
simplicity, we assume equilibrium between condensate and thermal cloud
within each well). As a result, the first
Josephson equation is modified to read
\be
\label{disip2}
\frac{d}{dt} N = \frac{E_J}{\hbar} \sin \varphi  +I_n,
\ee
or, for small fluctuations [using (2) and (\ref{Ohmic})],
\beq
\label{damped}
\ddot{\varphi}+GE_C\dot{\varphi}+\omega_{JP}^2\varphi=0.
\eeq
The classical harmonic oscillator described by eq. (\ref{damped}) is
underdamped if
\beq
\sigma \equiv \frac{1}{\hbar G }\sqrt{\frac{E_J}{E_C}} \gg 1.
\eeq
It only rests to calculate $G$ for a specific system. A simple estimate can
be made for the case
in which the chemical potential lies near the top of 
the barrier, at a distance $V_1 \sim\hbar\omega_0/2$, so that $S \sim 1-5$.
In 
this regime, we can still expect the WKB formula for $E_J$ derived 
above to
yield a reasonable approximation. However, if
$k_BT\gg\hbar\omega_0/2\sim V_1$, the thermal cloud lies mostly 
above
the barrier and a radically different approach is needed to study 
its
transport properties. For simplicity, we introduce the drastic
approximation that particles impinging on the barrier with energy 
$E$
are transmitted with probability one if $E>V_0$, and zero if 
$E<V_0$.
Then, the flow of normal atoms due to a fluctuation in $\mu$ is
only limited by the ``contact resistance", a concept taken from
ballistic transport in nanostructures \cite{imry86}. We may write  $G\equiv
N_{\rm 
ch}/h$,
where $N_{\rm ch}$ is an effective number of
available transmissive channels.  Within a continuum 
approximation,
and assuming that only transverse channels with a minimum energy
between $\mu_0$ and $\mu_0+k_BT$ are populated, we find (taking 
$\mu
\ll k_BT$) $N_{\rm
ch}=mA_nk_BT/2\pi\hbar^2$, where $A_n$ is a mean transverse
contact area seen by normal bosonic particles \cite{zapa98}.
Approximating \cite{gior97} $A_n \sim 2\pi k_BT/m\omega_0^2$, we
find $N_{\rm ch} \sim (k_BT/\hbar\omega_0)^2$, and hence 
\be
\label{estim}
\sigma
\sim 2 \pi N_0^{7/15} (\hbar \omega_0/k_BT)^2 e^{-S/2}.
\ee
Interestingly, the same result is formally obtained in the high barrier
limit, where normal atom tunneling also lies in the WKB regime
\cite{zapa98}. This gives additional confidence in the adequacy of the
estimate (\ref{estim}).
Taking $k_BT=10\hbar \omega_0$, we find $\sigma \sim 0.38$ for
$N \sim 10^4$, and 3.3 for $N \sim 10^6$, if $S \sim 5$. The 
equivalent numbers for $S \sim 1$ are 2.8 and 24. 
The conclusion is that coherent Josephson dynamics can be 
observed in current atomic Bose condensates if the barrier is low. 
Underdamped dynamics can be further favored by decreasing $T$ and 
increasing $N$.

\section{INTERNAL JOSEPHSON EFFECT}

It is possible to couple different hyperfine states of an atom (say $|Fm>$
and $|F'm'>$) through laser light. In particular, by applying a laser
pulse, one may force an atom initially in state $|Fm>$ to evolve in a
controllable way towards another state $|F'm'>$. If such a pulse is applied
to a macroscopic condensate, then it is the whole ensemble of atoms what
evolves coherently. Thus it is possible to prepare a condensate in which
each atom is in the same coherent superposition of the two states. Such an
experiment has been realized by Hall {\it et al.} \cite{hall98} with the
$|1,-1>$ and $|2,1>$ states of $^{87}$Rb (hereafter,  $|A>$ and $|B>$). A
natural variant of this experiment consists in using laser light which also
induces the opposite rotation ({\it e.g.} linearly polarized light), in
such a way that the two states are coherently connected in both directions,
very much like Cooper pairs in a Josephson junction can tunnel from left to
right and vice versa. In such a setup, it will be possible to study the
novel and, potentially, extremely rich ``internal" Josephson effect, which
bears some analogies with the physics of superfluid $^3$He-A, as noted by
Leggett \cite{legg98}.

The analysis of the spatial (external) Josephson effect given in section
III is based on the knowledge of $V_{\rm ext}(\br)$, where $\br$ we can be
varied continuosly between wells. A similar study of the internal Josephson
effect would not be practical. Fortunately, all is really needed is the
value of the matrix element connecting states $|A>$ and $|B>$, which can be
known by other means. Then, it is most convenient to employ a two-site
description of the Josephson link. We write \cite{milb97,vill97}
\beq
\label{twosite}
H=-\frac{\hbar\omega_R}{2}(a^{\dagger}b+ b^{\dagger}a)
+\frac{u}{2}[(a^{\dagger}a)^2+ (b^{\dagger}b)^2],
\eeq
where $a^{\dagger}$($b^{\dagger}$) creates one atom in state $|A>$($|B>$),
$u$ accounts for the interactions, and $\omega_R$ is the Rabi frequency.
For simplicity, we assume that the two atomic energies, as well as the
intraspecies interactions, are equal, and neglect the interspecies
interaction. Particle number conservation requires $a^{\dagger}a+
b^{\dagger}b=N_T \equiv 2N_0$.

To establish the connection with the pendulum Hamiltonian, we note that,
since particle eigenstates admit a representation
$\Phi_N(\varphi)=(2\pi)^{-1/2}\exp(-iN\varphi)$, we may write for (3)
\beq
\label{matrix1}
<N+1|H|N>=-E_J/2.
\eeq
A similar analysis for the two-site Hamiltonian (\ref{twosite}) yields
\beqa
\label{matrix2}
<N+1|H|N> &=& -(\hbar\omega_R/2) <N_A+1,N_B-1|a^{\dagger}b|N_A,N_B>
\nonumber \\
&=& -(\hbar\omega_R/2) \sqrt{(N_A+1)N_B}.
\eeqa
Noting that $N_{A,B}=N_0 \pm N$, and identifying (\ref{matrix1}) and
(\ref{matrix2}), we arrive at
\beq
\label{EJN}
E_J(N)=\hbar\omega_R \left[ N_0(N_0+1)-N(N+1) \right]^{1/2}.
\eeq
In the limit $1 \ll N \ll N_0$, eq. (\ref{EJN}) becomes
\beq
E_J=N_0\hbar\omega_R,
\eeq
which is a clear manifestation of the phenomenon of Bosonic amplification.
The identification of (\ref{twosite}) with the pendulum Hamiltonian is
completed 
by noting that (apart from constant energy shifts) the interaction term can
be written $uN^2$, {\it i.e.} $u=E_C/2$.

\section{CROSSOVER FROM JOSEPHSON TO RABI DYNAMICS}

A clear limitation of the standard pendulum Hamiltonian (3) is that, in the 
non-interacting limit $E_C \to 0$, the dynamics is suppressed; in
particular, the Josephson plasma frequency (\ref{WJP}) for small collective
oscillations vanishes. This contradicts our physical notion that
noninteracting atoms should indeed exhibit some dynamics. For instance,
under the effect of Hamiltonian (\ref{twosite}), an atom initially prepared
in state $|A>$ will undergo Rabi oscillations between states $|A>$  and
$|B>$ with frequency $\omega_R$. This effect is clearly beyond the scope of
a rigid (with $N$-independent $E_J$) pendulum model such as (3). However,
it should be describable by (\ref{twosite}), which has a well-defined
non-interacting limit. 
Thus, it seems of interest to develop a unified description which views
collective Josephson behavior and single atom Rabi oscillations as
particular cases of a more general dynamics. To that end, let us rewrite
(\ref{EJN}) in a semiclassical form,
\beq
\label{EJNsemi}
E_J(N)= \hbar\omega_R\sqrt{N_0^2-N^2}.
\eeq
The classical dynamics of the resulting non-rigid pendulum Hamiltonian,
\beq
\label{Hsemi}
H=-N_0\hbar\omega_R\sqrt{1-N^2/N_0^2}\cos\varphi+(E_C/2)N^2,
\eeq
has been analyzed by Smerzi {\it et al.} \cite{smer97}. In the limit of
small $\varphi$ and $N$, (\ref{Hsemi}) acquires the harmonic form
\beq
H=N_0\hbar\omega_R\frac{\varphi^2}{2} + \left( E_C+\frac{\hbar\omega_R}{N_0}
\right)\frac{N^2}{2},
\eeq
with a natural oscillation frequency
\beqa
\omega^2 &=& \omega_{JP}^2+\omega_R^2 \nonumber \\
\label{oscillation}
&=& (N_0E_C/\hbar)\omega_R+\omega_R^2.
\eeqa
It is thus clear that, for a given interaction strength, the double BEC
system can be driven continuously from the Josephson to the Rabi regime by
varying $\omega_R$, something feasible with current laser technology.
Noting that $E_C \sim \mu_0/N_0$, the Josephson limit corresponds to $\mu_0
\gg \omega_R$.

The difference between Josephson and Rabi dynamics is more dramatic when
one considers large values of $N$, comparable to $N_0$. When $E_C=0$, it is
easy to see that (\ref{Hsemi}) always yields oscillations of $N(t)$ around
zero with frequency $\omega_R$. By contrast, when $E_C \neq 0$, the
frequency depends on the initial value $N(0)$ and, when the ratio $\Lambda
\equiv N_0E_C/\hbar\omega_R$ exceeds a critical value $\Lambda_c$, the
oscillations average to a nonzero number \cite{milb97,smer97}. This number
$\Lambda_c$ depends on $N(0)$ and $\varphi(0)$. In particular,
$\Lambda_c=2$ when $N(0)$ takes its maximum value $N_0$, and $\Lambda_c
\rightarrow \infty$ when $N(0) \rightarrow 0$, {\it i.e.} in the small
amplitude limit, there are always oscillations of zero average and
frequency given by (\ref{oscillation}).
The phenomenon by which, due to macroscopic quantum coherence, it is
possible (for $\Lambda > \Lambda_c$) to create a large, long lived particle
number imbalance between two connected BEC's, is called quantum
self-trapping \cite{milb97,smer97}, and it is equivalent to the nonzero
average angular momentum  of a mechanical pendulum undergoing complete
turns. Like the pendulum motion, the average particle imbalance can decay
because of dissipation. 
Quantum self-trapping cannot be realized in superconducting Josephson
junctions because, the chemical potential differences would have to be
greater than the superconducting gap \cite{smer97}, thus allowing
quasiparticles to intervene and complicate the dynamics. This is not an
important limitation in the case of BEC's. There exists a close and
well-studied analog of quantum self-trapping in the longitudinal NMR of
$^3$He-A \cite{legg98}.

Eq. (\ref{Hsemi}) describes in a unified way Josephson and Rabi ($E_C=0$)
dynamics. This is analogous to how eq. (\ref{GPH}) can yield both the
Gross-Pitaevskii and the Schr\"odinger ($g=0$) equations for the wave
function of a many boson system.
The crossover between collective Josephson and individual Rabi dynamics
cannot be studied in superconductors and superfluids, because there
interactions are never completely negligible. It is a nice feature of 
Bose-Einstein condensation that it will allow us to study the crossover
between these two qualitatively different dynamical regimes in an elegant
fashion.

\section{PHASE DYNAMICS OF SEPARATE CONDENSATES}

The class of Josephson links describable by the harmonic approximation
(\ref{HO}) includes many of the foreseeable 
connections that will be realized between BEC's.
At equilibrium, the phase for such systems is narrowly peaked around
$\varphi=0$.
Being canonically conjugate variables, $N$ and $\varphi$ satisfy the
uncertainty relations $\Delta N \Delta \varphi \ge 1/2$. In general, their
m.r.s. values are
\beq
\Delta\varphi=\left(E_i/E_J  \right)^{1/2} \;\;\;
\Delta N=\left(E_i/E_C \right)^{1/2},
\eeq
being $E_i=(\hbar\omega_{JP}/2) \coth (\hbar\omega_{JP}/2k_BT)$ the average
energy of the oscillator. If $E_C$ and $k_BT \ll E_J$, we have
$\Delta\varphi \ll 1$.
It is then of interest to analyze the phase dynamics of two BEC's which,
having formed an equilibrium Josephson link, are suddenly disconnected
($E_J$ is made zero). Such a study may allow us to understand better the
robustness of the phase definition when the support of the Josephson
coupling is no longer present. One may naively think that two separate
BEC's have a well-defined relative particle number and that the phase is
necessarily random. However, if the starting point is that of a
well-defined phase, the situation is much less clear, since phase
randomization needs a finite time that may be sensitive to the details of
the environment. After all, we are all familiar with physical variables
(such as the position of macroscopic objects) which are permanently
well-defined and yet are not constants of motion. There are at least three
physical factors which favor this long term definition of non-conserved
quantities: (i) large inertia (small effective $E_C$), (ii) high potential
barriers in the effective potential landscape, and (iii) damping due to the
environment. For the phase of a Josephson junction, at least the first two
ingredients (and some times the three of them; see section IV) are present.
Here we wish to understand what is the fate of the phase when the second
factor (the Josephson coupling) is removed.

The randomization of the phase after suppression of the Josephson coupling
was investigated in refs. \cite{legg91} and \cite{sols94} for the case of
superconductors and superfluids. The possibility of performing 
interference experiments between BEC's \cite{andr97,hall98} has revived the
interest in the study of this \cite{barn96} and related problems
\cite{wrig96}. 

\subsection{Ballistic randomization}

In the absence of external perturbations, the evolution of the 
wave function
at $t>0$ 
is that of an initially narrow wave packet on a circle.
Assuming for the moment that $\varphi$ is an extended variable,
the phase uncertainty at long times is
\beq
\label{ball}
\Delta\varphi(t)\simeq \sqrt{E_iE_C}t/\hbar=\Delta N E_Ct/\hbar
\eeq
The ballistic time $\tau_B$ is defined as the time it takes 
$\Delta\varphi$
to be of order $2\pi$. We get
\beq
\tau_B=h/\sqrt{E_iE_C}.
\eeq
A typical number for superconductors can be $\tau_B \sim 10^{-8}$ s, while,
for superfluids, the equivalent time may be of the order of minutes or
hours, depending on whether the capacitive forces are due to
compressibility (fixed volume) or gravity \cite{legg91,sols94}. The
randomization of the relative phase between BEC's lies in an intermediate
time scale. Using (\ref{EC}), we find $\tau_B \sim 20-100$ ms (with $N \sim
10^4-10^6$) for $T=100$ nK. 

\subsection{Recurrent dynamics}

The phase is actually an angular variable for which $0$ and $2\pi$
are to be identified and this implies that
$N$ has integer eigenvalues. When $E_J=0$, 
the wave function evolves as 
\beq
\label{evol}
\Phi(\varphi,t)=<\varphi|\Phi(t)> = 
\sum_N a_N \Phi_N(\varphi) \exp (-iE_CN^2t/2\hbar),
\eeq
where $a_N=<N|\Phi(0)>$. From (\ref{evol}), we notice the existence of a 
recurrence time \cite{sols94}
\beq
\tau_R=2h/E_C,
\eeq
after which the wave packet recombines, 
$\Phi(\varphi,\tau_R)=\Phi(\varphi,0)$. The property $E_i\ll E_J$
implies $\tau_R \gg \tau_B$, so that the two time scales are 
well separated. This observation renders meaning  to the 
apparently
oversimplified mechanism of ballistic spreading, 
provided one considers times much smaller than $\tau_R$. 

A detailed analysis reveals the following picture \cite{sols95}:
An initially narrow wave packet begins
to broaden according to 
(\ref{ball}). After a time of order $\tau_B$,
the wave packet 
is almost uniformly spread around the circle, apart from fast 
oscillations. At a much later time
$\tau_R$, the wave packet recombines and broadens again
in a time scale $\tau_B$. 
A typical value may be
$\tau_R \sim 10^{-5}$ s for a superconductor, and an astronomically large
one for a superfluid. Again, BEC's lie somewhere in between: We find
$\tau_R \sim 30-500$ s for $N\sim 10^4-10^6$. It must be noted that
recurrence would only be observed in quite ideal conditions (very low
noise, good time resolution, 
etc.), perhaps not easily realizable.

\subsection{Diffusive randomization}

So far we have neglected the possible effect
of dissipation due to the environment. 
In refs. \cite{sols94,sols95}, a source of noise was studied
which, when really present,
can be expected to dominate the dynamics, namely,
the exchange of particles with a normal reservoir. In BEC's, such a
reservoir could be provided by the thermal cloud.

Following ref. \cite{sols94}, we base our study on a useful analogy which
can save us a good amount of work. If the contact between the condensate
and the reservoir is Ohmic, {\it i.e.} the rate of particle exchange
behaves as 
\beq
\delta\dot{N}=-\Gamma \delta\mu.
\eeq
where $\delta\mu$ is a possible chemical potential difference between
condensate and thermal cloud, then our problem is formally identical to
that of 
a Brownian particle (BP). 
The phase plays the role of the position and the capacitance is
analogous to the mass. The particle number $N$ is equivalent to 
the momentum $p$ (both are conserved quantities that are exchanged 
with the
bath), and the chemical potential difference $\mu \sim \dot{\varphi}$ 
is the
the analog of the particle velocity $v$ (the relative position of two BP's
behaves like another BP).
The friction coefficient $\eta$, defined as
$\dot{p}=-\eta v$ (where $v$ is a deviation of the velocity from its zero
equilibrium value), is equivalent to the coefficient $\Gamma$. 

Once the analogy with the BP 
has been established, we may borrow from the classical theory \cite{reif65} 
and 
write for the evolution of
the phase uncertainty
\beq
\label{dphit}
\Delta\varphi^2(t)=\frac{2k_BT}{\hbar^2\Gamma}\left[t-\frac{1}{\gamma}
(1-e^{-\gamma t})\right],
\eeq
where $\gamma=\Gamma E_C$.
At short times ($\gamma t \ll 1$), we reproduce the ballistic 
randomization
given by (\ref{ball}). By contrast, at long times ($\gamma t \gg 1$), the 
randomization
becomes diffusive,
\be
\label{diffu}
\Delta\varphi(t) \simeq \left(2k_BTt/\hbar^2\Gamma\right)^{1/2}. 
\eeq
Eq. (\ref{diffu}) is a classical result. However,
thanks to the work of ref. \cite{haki84} on a quantum BP
with Ohmic dissipation, we know that the diffusive law (\ref{diffu})
applies at long 
times
$t \gg \hbar/k_BT \gg \gamma^{-1}$,
provided $\Delta\varphi(0) \ll 1$

Analogously to the ballistic case, it is possible to define a time for
diffusive randomization
\beq
\tau_D = h^2\Gamma/2k_BT
\eeq
For superconductors, $\tau_D$ may be of the order of minutes, 
a surprisingly long time if we compare it with the estimates 
for $\tau_B$.
This is a very important effect.
The large disparity between the two time scales 
may be understood by noting that, while
the fast ballistic randomization is due to the 
large value of $E_C$ (small electrostatic capacitance), 
the diffusive dynamics (which takes 
over at a time $\gamma^{-1} \ll \tau_B$) is essentially 
independent of it
and is determined instead by the coupling to the normal reservoir.
We arrive at the apparently paradoxical result that, upon suppression of
the Josephson contact, dissipation may help to preserve phase coherence.

\subsection{Common vs. independent thermal clouds}

It is important to emphasize that the diffusive mechanism described above
only applies if the normal particle reservoir is {\it common} to both
condensates. 
In VII.C, we have considered a situation in which, although the two
condensates have been separated, the two thermal clouds remain connected to
the point of behaving as a single one. In superconductors, an equivalent
arrangement is not fundamentally difficult (the reservoir would be a normal
metal). In the case of BEC's, we may have a common thermal cloud when the
increase in the barrier height is large enough to render the two
condensates unconnected, but still sufficiently small to allow for an
essentially single thermal cloud. 

If the two reservoirs are independent, then, following the BP analogy (in
which the finite masses of each of the separate baths would cause a
ballistic spread of the uncertainty in the relative position), the finite
effective value of the total $E_C$ for each one of the condensate plus
reservoir system will give rise to a ballistic randomization of the
relative phase. This seems to be the case in interference experiments
between different hyperfine states \cite{hall98}. 

The crucial difference between ballistic and diffusive spreading of the
phase is that, in the former case, there is no restoring force for
fluctuations in $N$. Therefore, any initial deviation $\Delta N$ gives rise
to a permanent $\Delta \mu$ which causes a ballistic spread of the phase
[see (\ref{ball})]. On the contrary, if randomization is diffusive, an
initial departure from equilibrium is quickly compensated and, in the
average, $N(t)$ and $\mu(t)$ change sign many times, in such a way that
$\Delta\varphi(t)$ grows slowly [see (\ref{diffu})].

\section{CONCLUSIONS}

I have presented an overview of the physics of the Josephson effect betwee
Bose condensed systems, with emphasis on the recently achieved BEC's in
trapped alkali gases. I have mostly focussed on those physical phenomena
that are likely to be observed only (or more easily) in these novel
systems. So I have omitted the discussion of problems (such as {\it e.g.}
steady particle flow under the action of an alternating chemical potential)
which may be viewed as straightforward applications of well known Josephson
physics \cite{baro82}. I have tried to underline the potential richness of
the physics displayed by weakly connected BEC's. One may have an external
(spatial) and an internal (hyperfine) Josephson effect. It seems possible
to explore the crossover between collective Josephson behavior and
independent boson Rabi dynamics. Finally, the observation of fascinating
phenomena such as quantum self-trapping and macroscopic interference
between separate Bose condensates seems also within reach of the emerging
BEC technology. Everything indicates that the experimental and theoretical
study of the Josephson effect between 
Bose-Einstein condensates will lead us to the exploration of most exciting
new physics.

\acknowledgments

I wish to thank Anthony J. Leggett and Ivar Zapata for many stimulating
discussions. 
This work has been supported by the Direcci\'on General de Investigaci\'on
Cient\'{\i}fica y T\'ecnica under Grant No. PB96-0080-C02, and in part by
the National Science Foundation under Grant No. PHY94-07194. The
hospitality of the Institute for Theoretical Physics of the University of
California at Santa Barbara, where part of this work was done, is
gratefully acknowledged.


\begin{thebibliography}{1}

\bibitem{jose62} JOSEPHSON B.D., {\it Phys. Lett.}, {\bf 1} (1962) 251.

\bibitem{baro82} 
BARONE A. and PATERN{\`O} B., {\em Physics and applications of the 
Josephson effect} (John 
Wiley \& Sons, New York) 1982.

\bibitem{aven85} AVENEL O. and VAROQUAUX E., {\it Phys. Rev. Lett.}, {\bf 
55} (1985) 2704.

\bibitem{ande95} ANDERSON M.H., ENSHER J.R., MATTHEWS M.R.,
WIEMAN C.E. and CORNELL E.A., {\it Science}, {\bf 269} (1995) 198.

\bibitem{brad95} BRADLEY C.C, SACKETT C.A., TOLLETT J.J.,  
and HULET R.G., {\it Phys. Rev. Lett.}, {\bf 75} (1995) 1786.

\bibitem{davi95} DAVIS K.B., MEWES M.O., ANDREWS M.R., VAN DRUTEN N.J.,
DURFEE D.S., KURN D.M., and KETTERLE W., {\it Phys. Rev. Lett.}, {\bf 75}
(1995) 3969.

\bibitem{mewe96} MEWES M.O., ANDREWS M.R., VAN DRUTEN N.J., KURN D.M.,
DURFEE D.S., TOWNSEND C.G., and KETTERLE W., {\it Phys. Rev. Lett.}, {\bf 
77} (1996) 416; {\bf 77} (1996) 988.

\bibitem{feyn63} FEYNMAN R.P., LEIGHTON R.B., and SANDS M., 
{\it The Feynman Lectures on Physics} (Addison-Wesley Pub. Co., 
Reading) 1963.

\bibitem{baym96} BAYM G. and PETHICK C.J., {\it Phys. Rev. Lett.}, {\bf 
76} (1996) 6.

\bibitem{pita61} PITAEVSKII L.P., {\it Zh. Eksp. Teor. Fiz.}, {\bf 40}, 
(1961) 646
[{\it Sov. Phys. JETP}, {\bf 13} (1961) 451]; GROSS E.P., {\it Nuovo
Cimento}, {\bf 20} (1961) 454; {\it J. Math. Phys.}, {\bf 4} (1963) 195.

\bibitem{gior97} GIORGINI S., PITAEVSKII L.P., and STRINGARI S., 
{\it Phys. Rev. Lett.}, {\bf 78} (1997) 3987.

\bibitem{ande84} ANDERSON P.W.,
{\it Basic Notions of Condensed Matter Physics}, (Benjamin-Cummings,
Menlo Park) 1984.

\bibitem{zapa98} ZAPATA I., SOLS F., and LEGGETT A.J., {\it Phys. Rev. A},
{\bf 57} (1998) R28.

\bibitem{stri96} STRINGARI S., {\it Phys. Rev. Lett.}, {\bf 77} (1996) 2360.

\bibitem{edwa96} EDWARDS M., RUPRECHT P.A., BURNETT K., DODD R.J.,
and
CLARK C.W., {\it Phys. Rev. Lett.}, {\bf 77} (1996) 1671.

\bibitem{dalf96} DALFOVO F., PITAEVSKII L.P. and STRINGARI S., {\it Phys.
Rev. A},
{\bf 54} (1996) 4213.

\bibitem{gior96} GIORGINI S., PITAEVSKII L.P., and STRINGARI S., {\it Phys.
Rev. A}, {\bf 54} (1996) R4633.

\bibitem{imry86} IMRY Y., in {\it Directions in Condensed Matter}, edited by
G. GRINSTEIN and E. MAZENKO (World Scientific, Singapore) 1986.

\bibitem{hall98} HALL D.S., MATTHEWS M.R., WIEMAN C.E., and CORNELL E.A., 
cond-mat/9805327.

\bibitem{legg98} LEGGETT A.J., {\it Proc. ICAP XVI}, edited by W.E. BAYLIS,
AIP Press, to be published.

\bibitem{milb97} MILBURN G.J., CORNEY J., WRIGHT E.M., and WALLS D.F., {\it
Phys. Rev. A}, {\bf 55} (1997) 4318.

\bibitem{vill97}
VILLAIN P., LEWENSTEIN M., DUM R., CASTIN Y., YOU L., IMAMOGLU A., and
KENNEDY T.A.B., {\it J. Mod. Opt.}, {\bf 44} (1997) 1775.

\bibitem{smer97} SMERZI A., FANTONI S., GIOVANAZZI S., and SHENOY S.R.,
{\it Phys. Rev. Lett.}, {\bf 79} (1997) 4950.

\bibitem{legg91} LEGGETT A.J. and SOLS F., 
{\it Found. Phys.}, {\bf 21} (1991) 353.

\bibitem{sols94} SOLS F., {\it Physica B}, {\bf 194-196} (1994) 1389.

\bibitem{andr97} ANDREWS M.R., TOWNSEND C.G., MIESSNER H.J., DURFEE M.S.,
KURN D.M., and KETTERLE W., {\it Science}, {\bf 275} (1997) 637.

\bibitem{barn96} 
BARNETT S.M., BURNETT K., and VACCARO J.A., {\it J. Res. Natl. Inst. Stand.
Technol.}, {\bf 101} (1996) 593;
IMAMOGLU A., LEWENSTEIN M., and YOU L., {\it Phys. Rev. Lett.}, {\bf 78}
(1997) 2511;
JAVANAINEN J. and WILKENS M., {\it Phys. Rev. Lett.}, {\bf 78} (1997) 4675,
and {\it Phys. Rev. Lett.}, {\bf 81} (1998) 1345; LEGGETT A.J. and SOLS F.,
{\it Phys. Rev. Lett.}, {\bf 81} (1998) 1344.

\bibitem{wrig96} 
JAVANAINEN J. and YOO S.M., {\it Phys. Rev. Lett.}, {\bf 76} (1996) 161;
WRIGHT E.M., WALLS D.F., and GARRISON J.C., {\it Phys. Rev. Lett.}, {\bf
77} (1996) 2158;
LEWENSTEIN M. and YOU M., {\it Phys. Rev. Lett.}, {\bf 77} (1996) 3489;
NARASCHEWSKI M., WALLIS H., SCHENZLE A., CIRAC J.I., and ZOLLER P., {\it
Phys. Rev. A}, {\bf 54} (1996) 2185;
CIRAC J.I., GARDINER C.W., NARASCHEWSKI M., and ZOLLER P., {\it Phys. Rev.
A}, {\bf 54} (1996) R3714;
WALLIS H., R\"OHRL A., NARASCHEWSKI M., SCHENZLE A., and MIESSNER H.J.,
{\it J. Mod. Opt.}, {\bf 44} (1997) 1751;
CASTIN Y. and DALIBARD J., {\it Phys. Rev. A}, {\bf 55} (1997) 4330;

\bibitem{sols95} SOLS F. and HEGSTROM R.A., in {\it Fundamental Problems in
Quantum Physics}, edited by M. FERRERO and A. VAN DER MERWE (Kluwer,
Dordrecht) 1995.

\bibitem{reif65} REIF F., {\it Foundations of Statistical and 
Thermal
Physics} (McGraw-Hill, New York) 1965.

\bibitem{haki84} HAKIM V. and AMBEGAOKAR V., {\it Phys. Rev. B}, 
{\bf 32} (1984) 423.

\end{thebibliography}
\end{document}